\documentclass[aip,jcp,showpacs,amsmath,amssymb,preprint,superscriptaddress]{revtex4-1}
\usepackage{graphicx}
\usepackage{subfigure}
\usepackage{grffile}
\usepackage{bm}
\usepackage{mhchem}
\usepackage{color}
\usepackage{hyperref}
\usepackage{natbib}

\newcommand{\1}{\mbox{\bf 1}}
\begin{document}

\title{First-principles generalized gradient approximation (GGA)+${\bf
    U^d}$+${\bf U^p}$ studies of electronic structures and optical
  properties in cubic HfO2}

\author{Jinping Li}
\email{lijinping@hit.edu.cn; jinping@yukawa.kyoto-u.ac.jp}
\affiliation{Center for Composite Materials, Harbin Institute of
  Technology, Harbin 150080, China}
\affiliation{Yukawa Institute for Theoretical Physics, Kyoto
  University, Kyoto, 606-8502, Japan}

\author{Songhe Meng}
\affiliation{Center for Composite Materials, Harbin Institute of
  Technology, Harbin 150080, China}
    
\author{Lingling Li}
\affiliation{Center for Composite Materials, Harbin Institute of
  Technology, Harbin 150080, China}
  
\author{Hantao Lu}
\affiliation{Yukawa Institute for Theoretical Physics, Kyoto
  University, Kyoto, 606-8502, Japan}

\author{Takami Tohyama}
\affiliation{Yukawa Institute for Theoretical Physics, Kyoto
  University, Kyoto, 606-8502, Japan}

\date{\today}

\begin{abstract}

  The electronic structures and optical properties of cubic \ce{HfO2}
  are calculated by generalized gradient approximation (GGA)+$U$
  approach. Without on-site Coulomb interactions, the band gap of
  cubic \ce{HfO2} is $2.92\,\text{eV}$, much lower than the
  experimental value ($5.7\,\text{eV}$). Introducing the Coulomb
  interactions of $5d$ orbitals on Hf atom ($U^d$) and of $2p$
  orbitals on O atom ($U^p$), we can reproduce the experimental value
  of the band gap. The calculated dielectric function of cubic
  \ce{HfO2} by the GGA+$U^d$+$U^p$ approach predicts the presence of a
  shoulder structure below the main peak of the absorption
  spectrum. These indicate that the GGA+$U^d$+$U^P$ approach is a
  convenient and powerful method to calculate and predict the
  electronic structures and optical properties of wide-gap optical
  materials.

\end{abstract}



\maketitle

\section{Introduction}

\ce{HfO2} has been widely studied both experimentally and
theoretically because of its excellent dielectric properties, wide
band gap, high melting point, {\em etc.}~\cite{Weir:1999,
  Cao:1998}. Moreover, \ce{HfO2} has been proved to be one of the most
promising high-dielectric-constant materials, due to its high
formation heat and nice thermodynamic stability when it contacts
indirectly with Si~\cite{He:2007}. It has also been used in
optical~\cite{Gilo:1999} and protective
coatings~\cite{Yamamoto:2002}. At atmospheric pressure, \ce{HfO2}
exists in three polymorphs when temperature changes~\cite{Terki:2008}:
monoclinic structure (P21/c, C2h 5) at low temperatures, tetragonal
(P42/nmc, D4h 15) at temperature higher than $2000\,\text{K}$, and
finally, with increasing temperature over $2870\,\text{K}$, the cubic
fluorite (Fm3m, Oh 5) becomes stable. However, by rapid quenching or
stabilizer addition, the cubic structure can be stabilized at low
temperatures, e.g., see Refs.~\onlinecite{Lee:2004, Shi:2010}.

The electronic and structural properties of \ce{HfO2} have been
theoretically studied by the first-principles band structure
calculation based on the density-functional theory (DFT). For example,
the structure, vibration and lattice dielectric properties of
\ce{HfO2} have been investigated in the frame of the local density
approximation (LDA) and generalized gradient approximation
(GGA)~\cite{Zhao:2002}. Based on the ABINIT package, the Born
effective charge tensors, phonon frequencies and dielectric
permittivity tensors in cubic and tetragonal phases are calculated by
LDA~\cite{Rignanese:2004}. The full potential linearized augmented
plane-wave method based on the WIEN2k code is also employed to study
the structural and electronic properties in the cubic
phase~\cite{Terki:2005}. Further, the spin-orbit effect is estimated
within the framework of LDA and GGA~\cite{Garcia:2005}. The resulting
band gap is about $3.65\,\text{eV}$, still smaller than the
experimental value ($5.7\,\text{eV}$)~\cite{He:2007}, indicating that
correlation effects in \ce{HfO2} cannot be ignored.

Jiang {\em et al.}~\cite{Jiang:2010} have studied the band structures
of \ce{HfO2} by the GW approximation, where the self-energy of a
many-body system is approximated in terms of the screened Coulomb
interaction (W) up to first order, and then correlation effects can be
addressed to some extent. The resulting band gap value is
$5.2\,\text{eV}$. Although the value is much closer to the
experimental one, the GW method is highly cost in terms of numerical
demands. Another technique to include correlation effects with
relatively less computational effort is the so-called LDA+$U$ or
GGA+$U$ approach, in which the correlation effect is incorporated
through the on-site Coulomb interaction $U$~\cite{Anisimov:1997}. As
an approximation to GW, GGA(LDA)+$U$ approach can reproduce
experimental data as accurately as the hybrid functional approach but
with much lower computational efforts~\cite{Jiang:2010b}. It can give
a qualitative improvement compared with the LDA, not only for
excited-state properties such as energy gaps, but also for
ground-state properties such as magnetic moments and interatomic
exchange parameters~\cite{Anisimov:1997}. Recently, a GGA+$U$ approach
for \ce{CeO2} has been performed, where not only the on-site $U$ for
Ce $4f$ electrons, but also for O $2p$ electrons, are
included~\cite{Plata:2012}. The choice of $U$ is not
unambiguous. Though there are attempts to extract it from the
first-principles calculations, it is nontrivial to determine its value
a priori. Hence, in practice, $U$ is often fitted to reproduce a
certain set of experimental data, such as band gaps and structural
properties.

In this paper, we use GGA and GGA+$U$ schemes formulated by Loschen
{\em et al.}~\cite{Loschen:2007}, to calculate the lattice parameters,
band structures, and optical properties of cubic \ce{HfO2}
(c-\ce{HfO2}). We find that the value of the band gap can be
reproduced by introducing the on-site Coulomb interactions of $5d$
orbitals on Hf atom ($U^d$) and of $2p$ orbitals on O atom ($U^p$). We
also notice that a shoulder structure below the main peak of the
imaginary part of the dielectric constant remains even when $U^d$ and
$U^p$ are introduced. This implies that the shoulder structure is
robust against the change of band structure due to the Coulomb
interactions. We thus expect that the shoulder will appear in
absorption measurements for pure c-\ce{HfO2}. These findings imply
that the GGA+$U^d$+$U^p$ approach is a convenient and powerful method
to calculate and predict the electronic structures and optical
properties of wide-gap optical materials.

\section{Computational Methodology}

Density functional theory calculations are performed with plane-wave
ultrasoft pseudopotential, by using GGA with Perdew-Burke-Ernzerhof
(PBE) functional and GGA+$U$ approach as implemented in the CASTEP
code (Cambridge Sequential Total Energy
Package)~\cite{Segall:2002}. The ionic cores are represented by
ultrasoft pseudopotentials for Hf and O atoms. For Hf atom, the
configuration is [Xe]$4f^{14}5d^{2}6s^{2}$, where the $5d^2$ and
$6s^2$ electrons are explicitly treated as valence electrons. For O
atom, the configuration is [He]$2s^22p^4$, and valence electrons
include $2s^2$ and $2p^4$. The plane-wave cut off energy is
$380\,\text{eV}$.  And the Brillouin-zone integration is performed
over the $24\times24\times24$ grid sizes using the Monkorst-Pack
method for cubic structure optimization. This set of parameters assure
the total energy convergence of $5.0\times10^{-6}\,\text{eV/atom}$,
the maximum force of $0.01\,\text{eV/\AA}$, the maximum stress of
$0.02\,\text{GPa}$ and the maximum displacement of
$5.0\times10^{-4}\,\text{\AA}$.

In the following sections, we firstly optimize the geometry structure
of c-\ce{HfO2} by the GGA method. Using the optimized structure, we
next introduce $U^d$ for Hf $5d$ orbitals and $U^p$ for O $2p$
orbitals. Comparing the numerical values of the band gap with the
experimental one, we obtain the best values of $U^d$ and $U^p$. The
electronic structures and optical properties of c-\ce{HfO2} are
calculated by means of GGA, without $U$ and with $U^d$+$U^p$,
respectively. Comparison with available experimental data is
presented.

\section{Results and Discussion}

\subsection{GGA Calculation}

The space group of c-\ce{HfO2} is Fm3m and the local symmetry is
O5h. Moreover, c-\ce{HfO2} is fully characterized by a single lattice
constant $a$. The GGA calculation of the perfect bulk c-\ce{HfO2} is
performed to determine an optimized $a$ in order to check the
applicability and accuracy of the ultrasoft pseudopotential. The
optimized $a$ is $0.526\,\text{nm}$, in good agreement with other
theoretical values~\cite{Zhao:2002, Rignanese:2004, Terki:2005,
  Liu:2009}. The deviation from experimental values
($0.508\,\text{nm}$~\cite{Wang:1992} and
$0.516\,\text{nm}$~\cite{Garcia:2005}) is $3.54$\% at most. Hereafter,
we use this optimized value of $a$.

\begin{figure}
\subfigure[]{\includegraphics[width=0.4\textwidth]{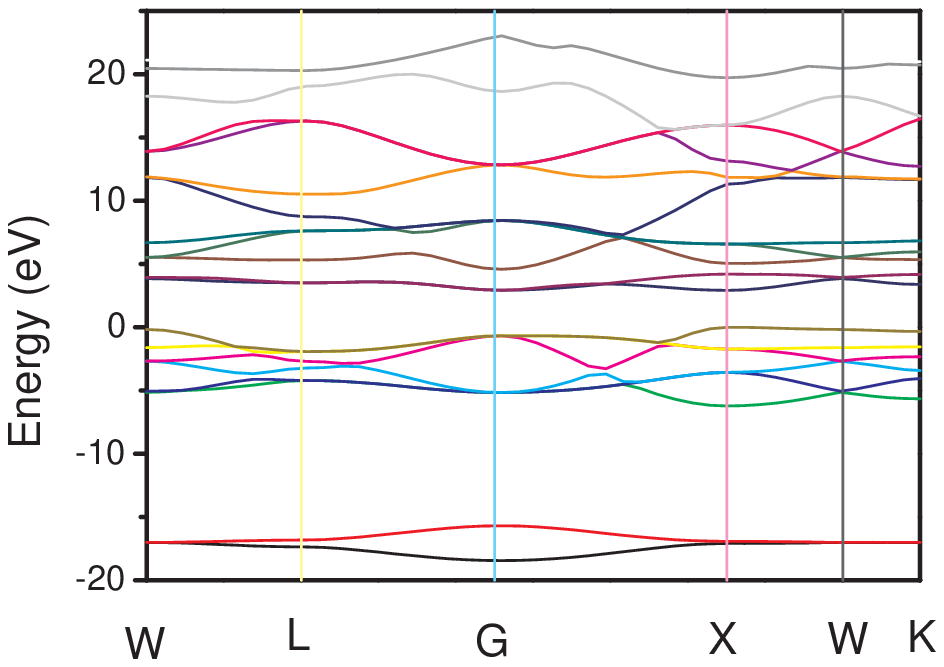}
\label{fig1a}}
\subfigure[]{\includegraphics[width=0.4\textwidth]{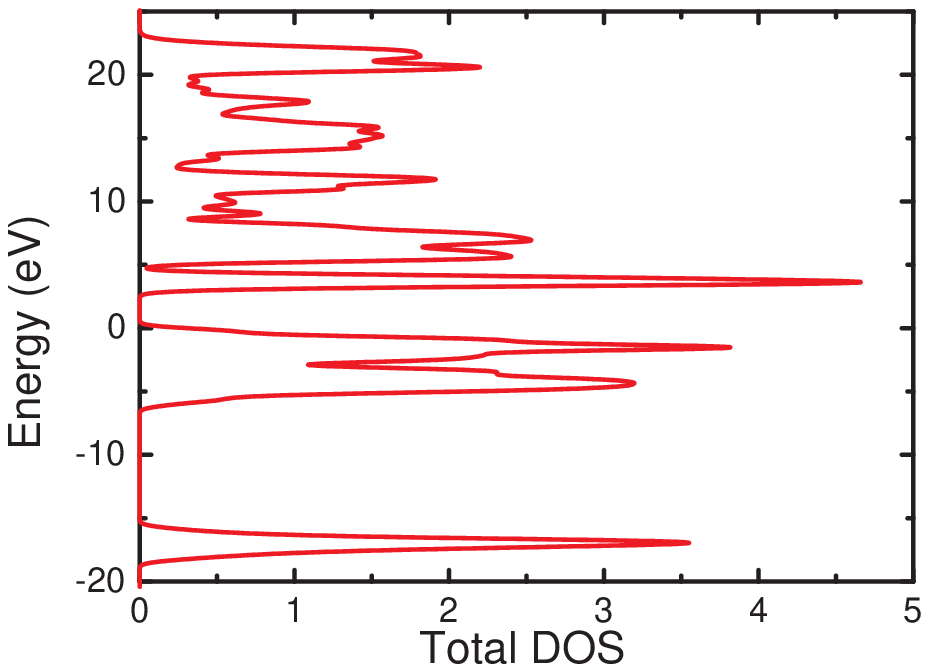}
\label{fig1b}}
\\
\subfigure[]{\includegraphics[width=0.4\textwidth]{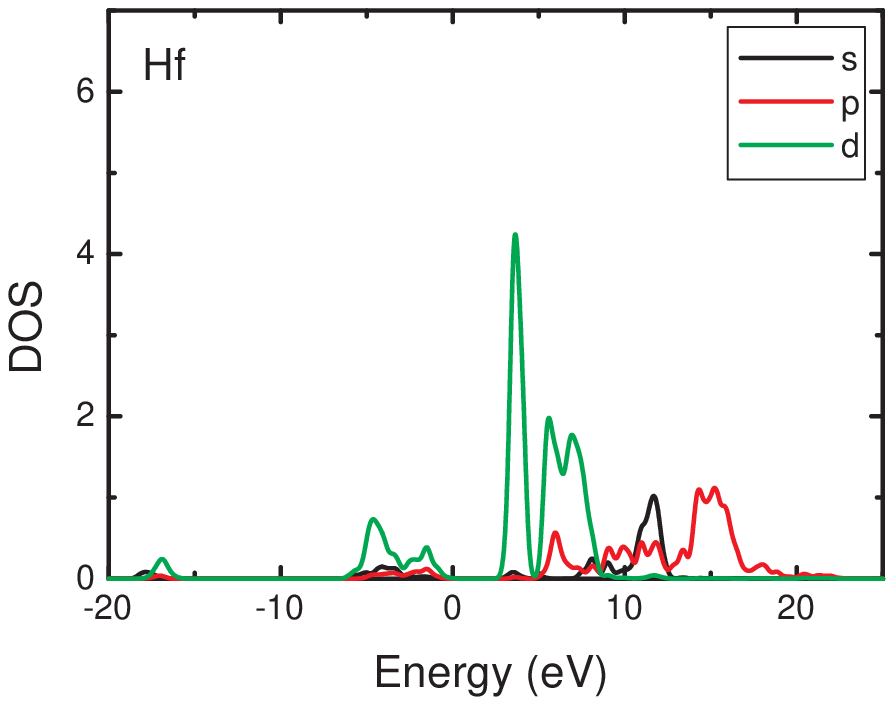}
\label{fig1c}}
\subfigure[]{\includegraphics[width=0.4\textwidth]{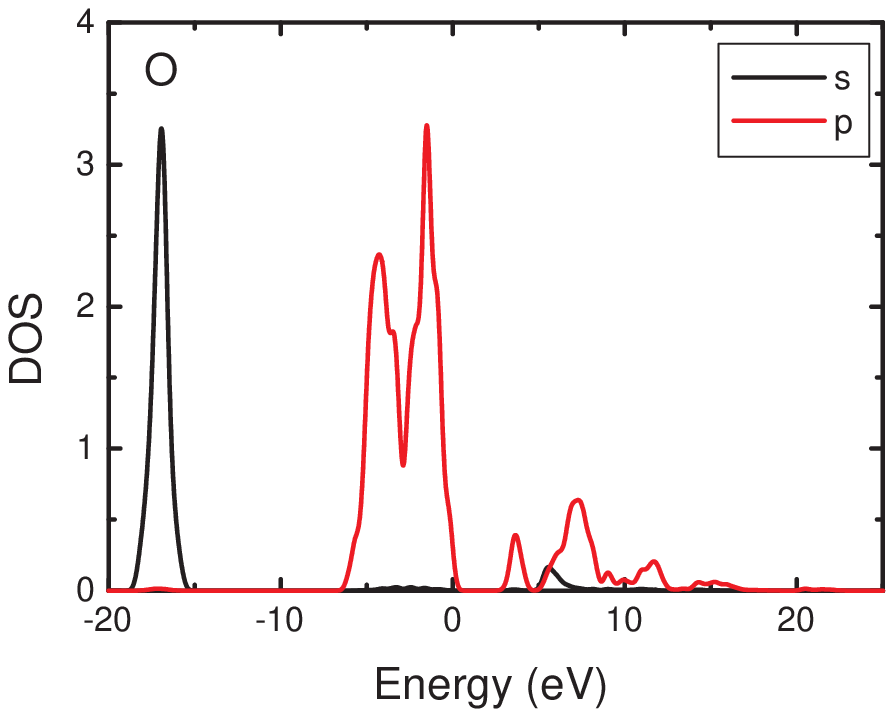}
\label{fig1d}}
\caption{The band structure and density of states (DOS) of c-\ce{HfO2}
  obtained by GGA. (a) Band structure; (b) total DOS; (c) partial DOS
  of Hf atom, and (d) partial DOS of O atom.}
\label{fig1}
\end{figure}

The band structure along high-symmetry directions of the Brillouin
zone and total density of states (DOS) of c-\ce{HfO2} are shown in
Fig.~\ref{fig1}. The band structure in Fig.~\ref{fig1a} shows a direct
band gap because the top of the valence bands and the bottom of the
conduction bands are found at the same X point. The value of the band
gap $E_g$ is around $2.92\,\text{eV}$, much smaller than the
experimental value $5.7\,\text{eV}$. This is due to the well-known
underestimate of conduction-band energies in {\em ab initio}
calculations: the DFT results often undervalue the energy of $5d$
orbitals of Hf atom, lowering the bottom level of conduction bands. As
a result, the band gap obtained by GGA is lower than the experimental
one.

The total DOS is presented in Fig.~\ref{fig1b}. There are two parts in
the valence band, namely, the lower region from $-18.9\,\text{eV}$ to
$-15.2\,\text{eV}$ and the upper region from $-6.6\,\text{eV}$ to
$0.1\,\text{eV}$. Figure~\ref{fig1d} shows that the lower valence
bands are predominantly composed of O $2s$, while the upper valence
bands consist of O $2p$ accompanied with hybridization with Hf $5d$ as
shown in Fig.~\ref{fig1c}. The conduction bands below $10\,\text{eV}$
are mostly composed of Hf $5d$ with some amount of O $2p$. The $6s$
and $5p$ orbitals of Hf atom also contribute to the conduction bands,
though their values are small compared with $5d$ states.

\begin{figure}
\includegraphics[width=0.5\textwidth, angle=-90]{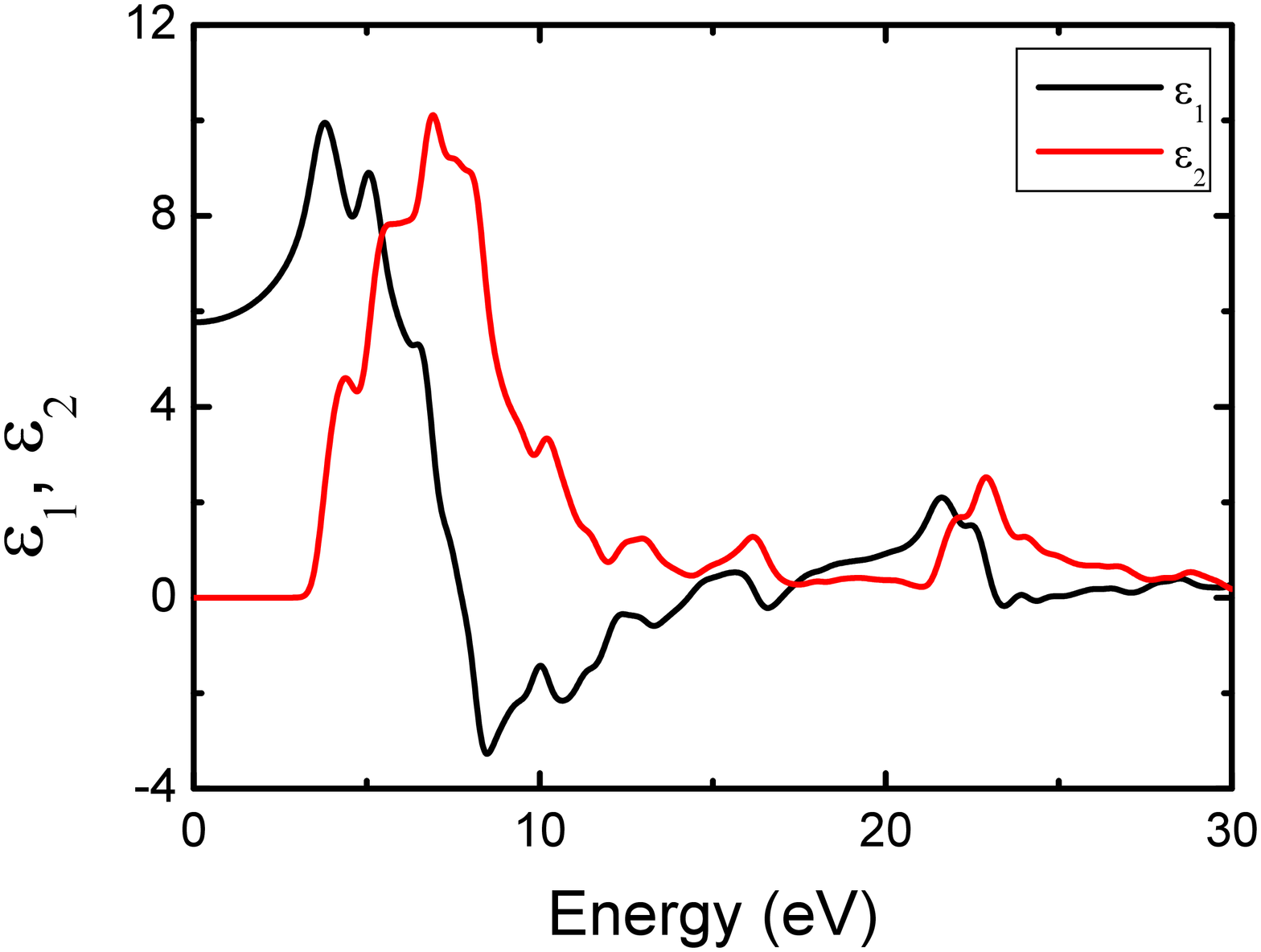}
\caption{Real and imaginary parts of dielectric function for
  c-\ce{HfO2} obtained by GGA.}
\label{fig2}
\end{figure}

Dielectric function is obtained by taking into account of inter-band
transitions. Figure~\ref{fig2} shows the complex dielectric function
as a function of photon energy. The real part of the dielectric
function $\epsilon_1$ in the low energy increases with photon energy
and gets maximum at $3.5\,\text{eV}$. Then it drops sharply with
photon energy to negative values and return back to positive
later. The imaginary part $\epsilon_2$ shows a maximum at
$7.5\,\text{eV}$ followed by two shoulder structures below the
energy. The two shoulders may be related to DOS in conduction bands at
$4\,\text{eV}$ and $7\,\text{eV}$ as shown in Fig.~\ref{fig1a}. Our
results of the dielectric function agree with other calculations for
c-\ce{HfO2}~\cite{He:2007, Park:2008}. However, the magnitude of
maximum value of $\epsilon_2$ ($\sim 10$) is larger than an available
experimental value ($\sim 8$) of \ce{HfO2}~\cite{Lim:2002}. This
disagreement will be resolved if the gap magnitude is reproduced by
introducing the on-site Coulomb interactions, as discussed below.

\subsection{GGA+${\bf U^d}$+${\bf U^p}$ Calculation}

\begin{figure}
\subfigure[]{\includegraphics[width=0.4\textwidth]{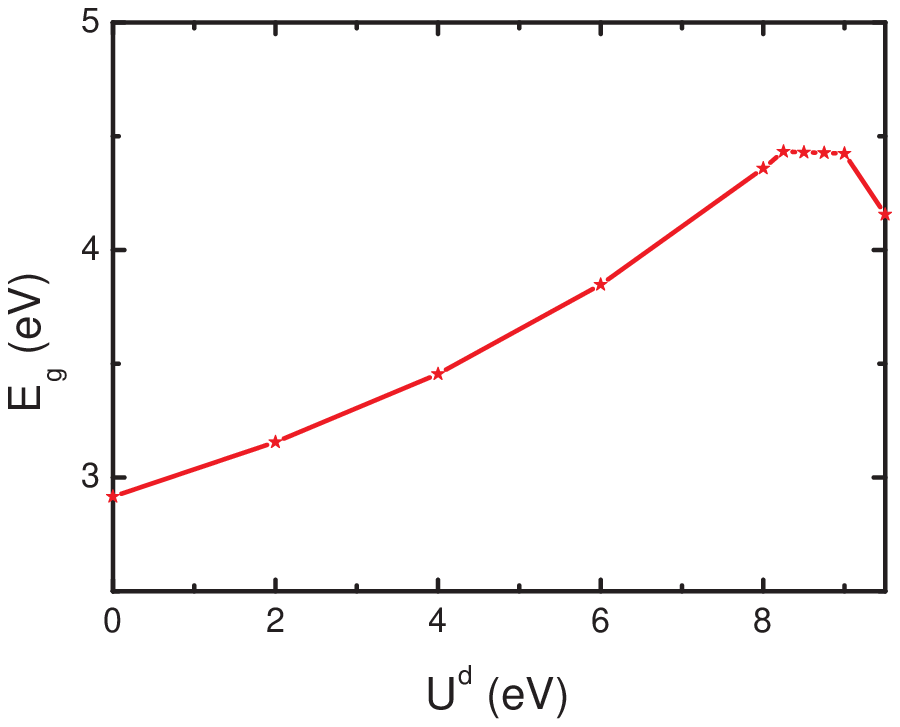}
\label{fig3a}}
\subfigure[]{\includegraphics[width=0.4\textwidth]{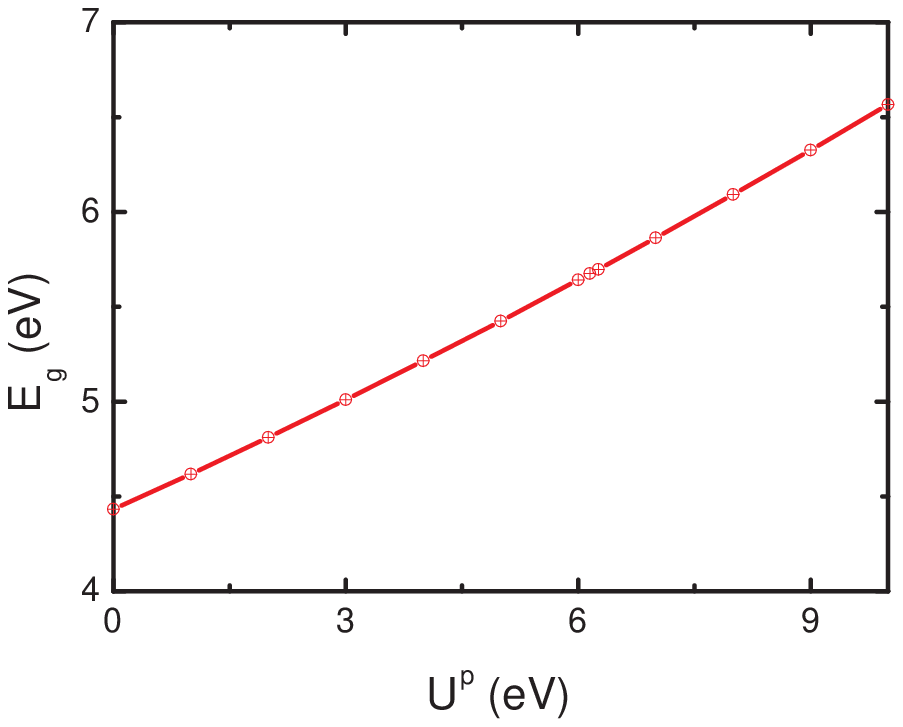}
\label{fig3b}}
\caption{Calculated band gap $E_g$ as a function of (a) $U^d$ and (b)
  $U^p$.}
\label{fig3}
\end{figure}

Using the optimized lattice parameter, $a=0.526\,\text{nm}$, we
calculate the band structure and DOS of c-\ce{HfO2} by changing $U^d$
for $5d$ orbitals of Hf atom. The band gap $E_g$ obtained from the
band structure is shown in Fig.~\ref{fig3a} as a function of $U^d$. It
can be seen that $E_g$ firstly increases, and then drops with
increasing $U^d$, showing a maximum value ($4.43\,\text{eV}$) at
$U^d=8.25\,\text{eV}$. The maximum value is smaller than the
experimental one ($5.7\,\text{eV}$). The saturation of $E_g$ with
$U^d$ may be related to the approach of $5d$ states toward $6s$ and
$5p$ states, though microscopic mechanism is not yet fully
understood. Next, we introduce $U^p$ for $2p$ orbital of O atom, while
keeping $U^d =8.25\,\text{eV}$. The result in Fig.~\ref{fig3b} shows
that $E_g$ monotonically increases with $U^p$.  When $U^d
=8.25\,\text{eV}$ and $U^p=6.25\,\text{eV}$, the calculated band gap
is $5.697\,\text{eV}$, coinciding with the experiment one.

There could of course exist different combinations of $U^d$ and $U^p$
which can reproduce the band gap. For example, $U^d=10\,\text{eV}$,
$U^p=6.35\,\text{eV}$, the resulting band gap is $5.702\,\text{eV}$,
with similar optical properties in the low-energy regime. Based on
physical consideration with reference to other numerical results,
here, we choose $U^d=8.25\,\text{eV}$, $U^p=6.25\,\text{eV}$ as a
typical representative and perform the GGA+$U$ calculations.

\begin{figure}
\subfigure[]{\includegraphics[width=0.4\textwidth]{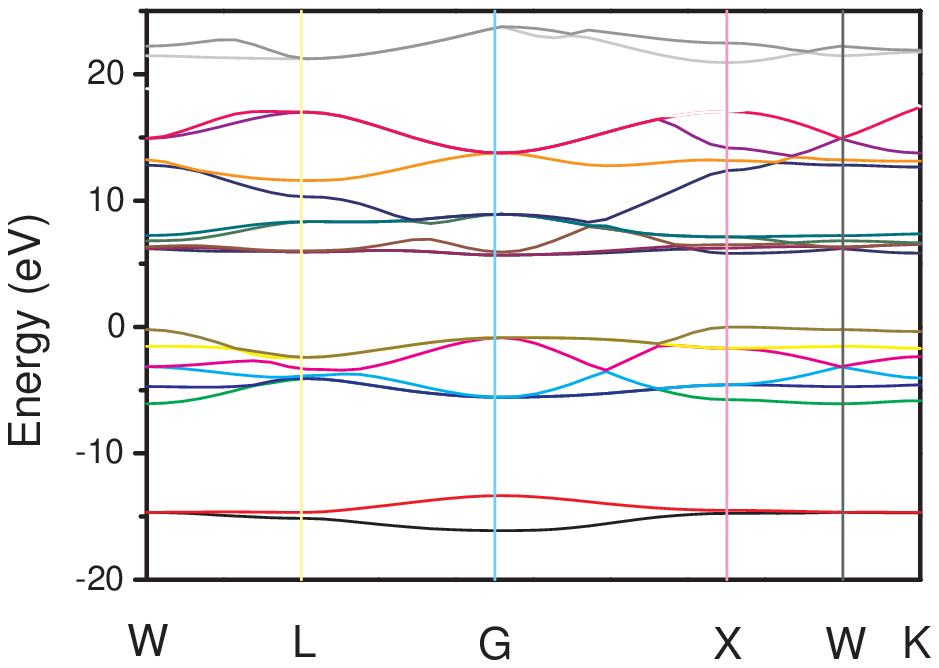}
\label{fig4a}}
\subfigure[]{\includegraphics[width=0.4\textwidth]{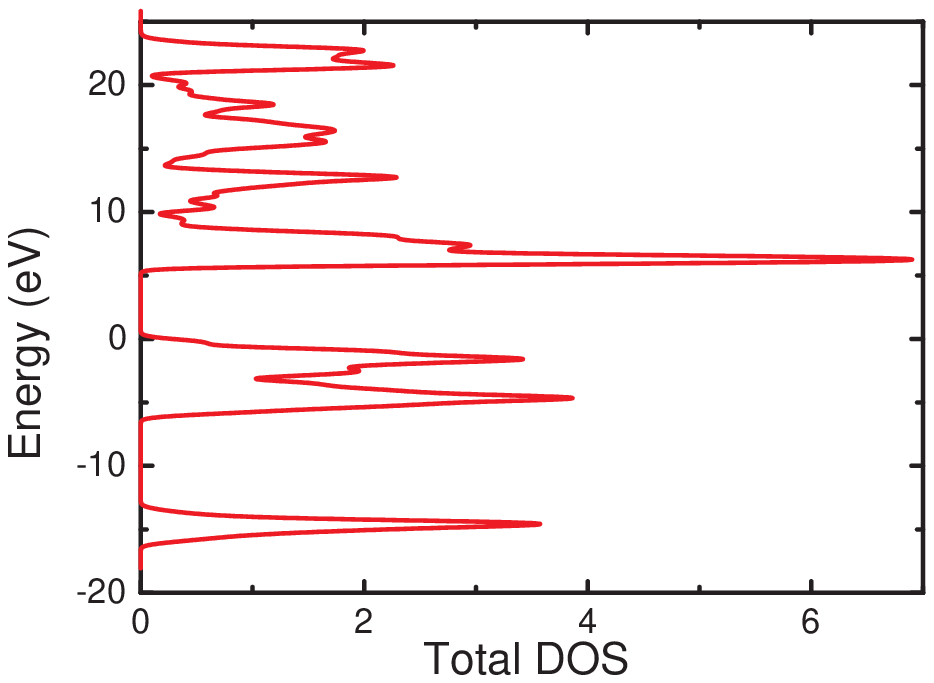}
\label{fig4b}}
\\
\subfigure[]{\includegraphics[width=0.4\textwidth]{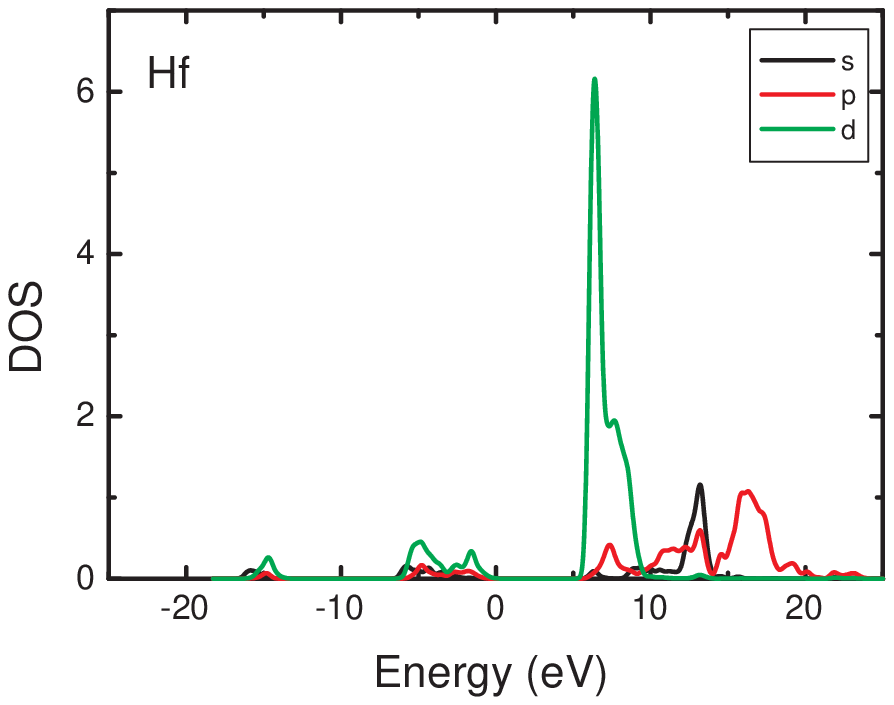}
\label{fig4c}}
\subfigure[]{\includegraphics[width=0.4\textwidth]{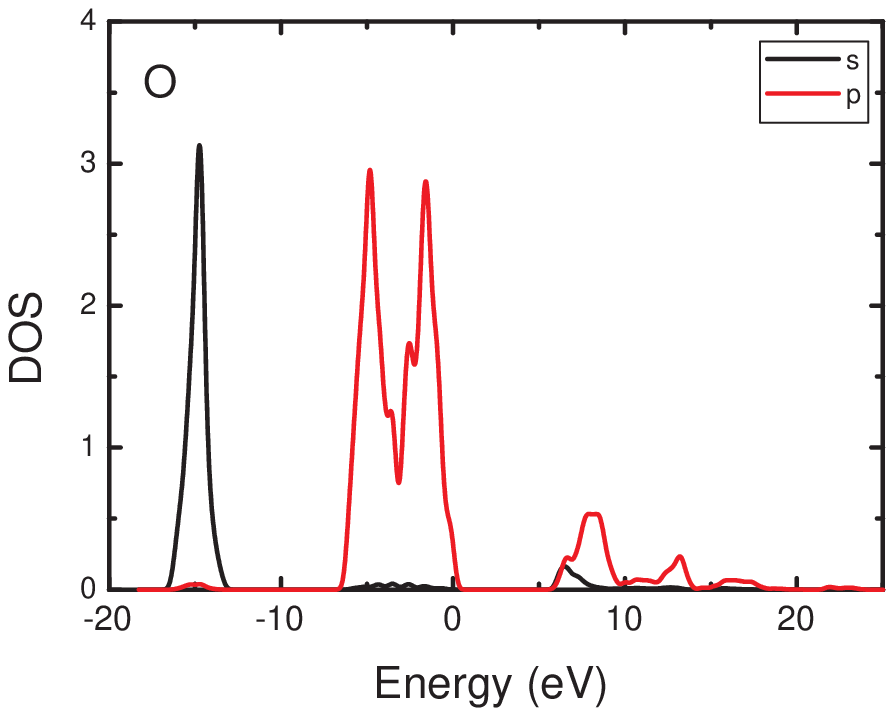}
\label{fig4d}}
\caption{The band structure and density of states (DOS) of c-\ce{HfO2}
  obtained by GGA + $U^d$ ($U^d=8.25\,\text{eV}$) + $U^p$
  ($U^p=6.25\,\text{eV}$). (a) Band structure; (b) total DOS; (c)
  partial DOS of Hf atom, and (d) partial DOS of O atom.}
\label{fig4}
\end{figure}

The resulting band structure is shown in Fig.~\ref{fig4a}. The bottom
of the conduction bands is moved to the G point. The band dispersions
near the bottom shift to higher energy with the increase of $U^d$ and
are reconstructed as compared with the dispersions in
Fig.~\ref{fig1a}. As a result, the separated DOS at $4\,\text{eV}$ and
$7\,\text{eV}$ in Fig.~\ref{fig1b} merges to one sharp structure at
$6\,\text{eV}$ in Fig.~\ref{fig4b}. It is clear from Fig.~\ref{fig4c}
that the reconstruction is caused by the $5d$ state of Hf. The O $2p$
states are also affected by the reconstruction through hybridization
effect as seen in Fig.~\ref{fig4d}. We note that the band gap
increases with increasing $U^p$ through such strong hybridization
effect.

\begin{figure}
\includegraphics[width=0.5\textwidth, angle=270]{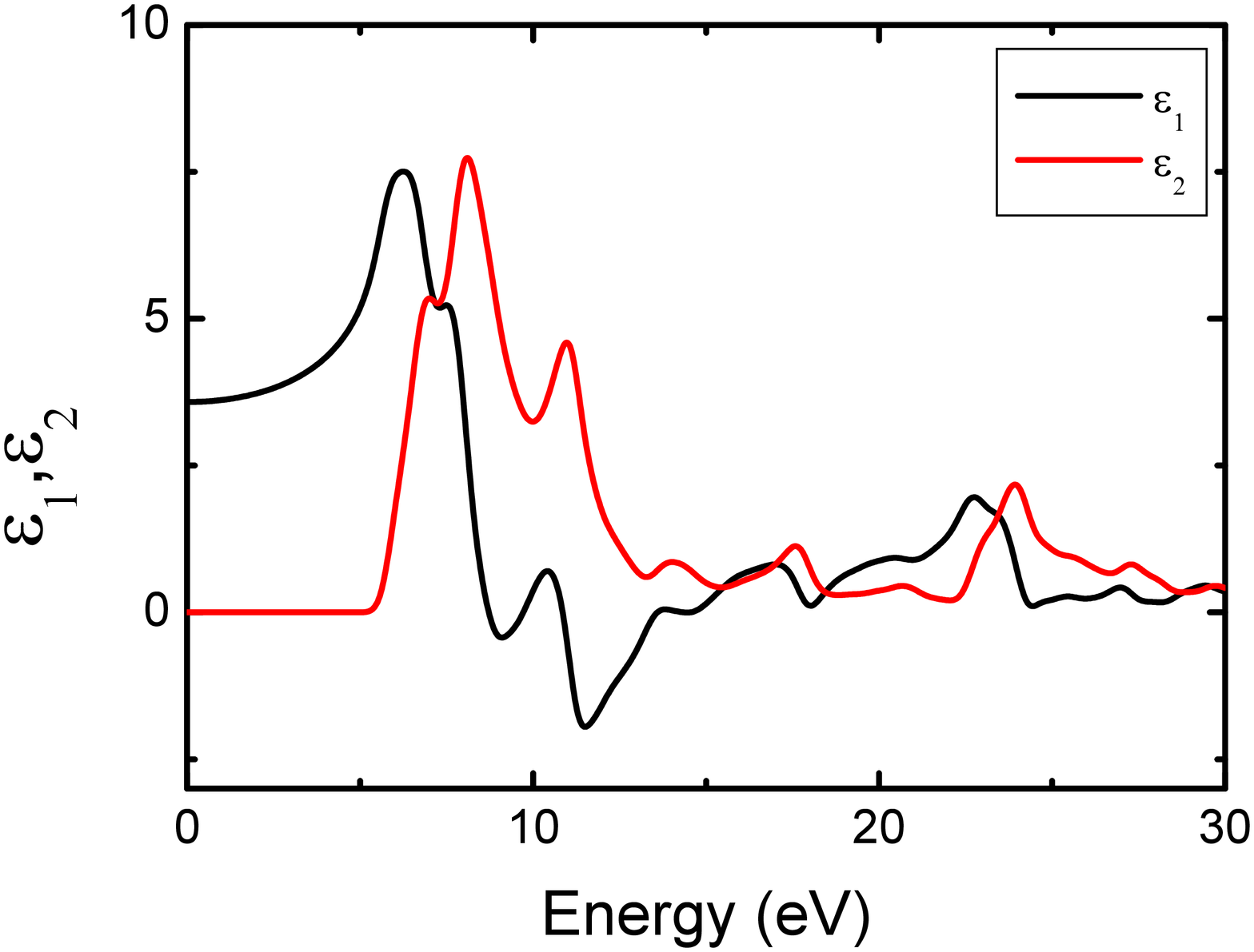}
\caption{Real and imaginary parts of dielectric function for
  c-\ce{HfO2} obtained by GGA+$U^d$+$U^p$.}
\label{fig5}
\end{figure}

Figure~\ref{fig5} shows the dielectric function. The real part
$\epsilon_1$ exhibits a maximum at $6.38\,\text{eV}$, corresponding to
band edge reflection peak at $5.0\,\text{eV}$ in the reflection
spectrum. The calculated static dielectric constant is $3.44$,
coinciding with the experimental value of monoclinic
\ce{HfO2}~\cite{He:2007, Koike:2006}. The imaginary part $\epsilon_2$
shows a maximum at $8\,\text{eV}$. The maximum value ($\sim 7.6$) is
very close to the experimental value ($\sim 8$)~\cite{Lim:2002}, in
contrast with the GGA case as mentioned before. This improvement comes
from the agreement of $E_g$ between the numerical and experimental
values. Another interesting observation in $\epsilon_2$ is the
presence of a shoulder structure below the peak. It is remarkable
that, even though the reconstruction of the band structure occurs near
the bottom of the conduction band, the shoulder structure remains as
is the case of the GGA calculation. This implies the robustness of the
shoulder structure in c-\ce{HfO2}. Therefore, we can expect that such
a shoulder structure appears in the absorption spectrum in pure
c-\ce{HfO2} samples.

Other optical properties can be computed from the complex dielectric
function~\cite{Wang:1992}. We have calculated the refractive index of
c-\ce{HfO2} by GGA+$U^d$+$U^p$, including refractive coefficient $n$
and extinction coefficient $k$. The results show that $n$ is $1.85$,
consistent with the experimental value of monoclinic \ce{HfO2},
$1.93$~\cite{He:2007}.

\section{Conclusions}

The structure optimization of c-\ce{HfO2} is performed by using
first-principles GGA and the resulting cell parameter of optimization
structure is $0.526\,\text{nm}$, consistent with the experimental
value and other theoretical results. However, the numerical value of
band gap is only $2.92\,\text{eV}$, much lower than the experimental
ones. The deviation can be resolved by introducing on-site Coulomb
interactions into $5d$ orbital of Hf atom ($U^d$) and $2p$ orbital of
O atom ($U^p$) at the same time. We obtain the best estimations of
$U^d$ and $U^p$ for c-\ce{HfO2}. The electron structure and optical
properties are calculated both by GGA and GGA + $U^d$ ($U^d
=8.25\,\text{eV}$)+ $U^p$ ($U^p=6.25\,\text{eV}$). We find that a
shoulder structure below the main peak of the imaginary part of the
dielectric constant remains even when $U^d$ and $U^p$ are
introduced. This implies that the shoulder structure is robust, and
we thus expect that the shoulder appears in the optical measurements
for pure c-\ce{HfO2} samples. These findings imply that the
GGA+$U^d$+$U^p$ approach is a powerful method to calculate and predict
the electronic structures and optical properties of wide-gap optical
materials.

\begin{acknowledgments}
  The authors thank Prof. Vladimir I. Anisimov and Dr. Alexey
  Shorikov, Institute of Metal Physics, Russian Academy of Sciences,
  for their valuable discussions.
\end{acknowledgments}

\providecommand{\noopsort}[1]{}\providecommand{\singleletter}[1]{#1}%

%\bibliography{cHfO}

\end{document}